\documentclass[a4paper,fleqn,usenatbib]{mnras}
\usepackage{newtxtext,newtxmath}

\usepackage[T1]{fontenc}
\usepackage{ae,aecompl}

\usepackage{graphicx}	
\usepackage{amsmath}	
\usepackage{amssymb}	
\title[Magnetic fields in bright points]{The magnetic properties of photospheric magnetic bright points with high resolution spectropolarimetry}
\vspace{-2.5cm}
\author[P. H. Keys et al.]{Peter H. Keys$^{1}$\thanks{E-mail: p.keys@qub.ac.uk},
Aaron Reid$^{1}$, 
Mihalis Mathioudakis$^{1}$, 
Sergiy Shelyag$^{2}$,\newauthor
Vasco M. J. Henriques$^{3,4}$, 
Rebecca L. Hewitt$^{1}$, 
Dario Del Moro$^{5}$, 
Shahin Jafarzadeh$^{3,4}$, \newauthor
David B. Jess$^{1,6}$, and  
Marco Stangalini$^{7,8}$
\\
$^{1}$Astrophysics Research Centre, School of Mathematics and Physics, Queen's University Belfast, Belfast, BT7 1NN, Northern Ireland, U.K.\\
$^{2}$School of Information Technology, Faculty of Science, Engineering and Built Environment, Deakin University, Geelong, Victoria, Australia \\
$^{3}$Institute of Theoretical Astrophysics, University of Oslo, P.O. Box 1029 Blindern, NO-0315 Oslo, Norway\\
$^{4}$Rosseland Centre for Solar Physics, University of Oslo, P.O. Box 1029 Blindern, N-0315 Oslo, Norway\\
$^{5}$Dipartimento di Fisica, Universit\`a degli Studi di Roma ``Tor Vergata'', via della Ricerca Scientifica 1, 00133 Roma, Italy\\
$^{6}$Department of Physics and Astronomy, California State University Northridge, Northridge, CA 91330, U.S.A\\
$^{7}$INAF-OAR National Institute for Astrophysics, Via Frascati 33, 00078 Monte Porzio Catone (RM), Italy\\
$^{8}$ ASI Agenzia Spaziale Italiana, Via del Politecnico snc, I-00133 Rome, Italy 
}

\vspace{-4.5cm}\date{Accepted 2019 June 13. Received 2019 June 13; in original form 2019 April 05}

\pubyear{2019}

\begin{document}
\label{firstpage}
\pagerange{\pageref{firstpage}--\pageref{lastpage}}
\maketitle

\vspace{-15cm}\begin{abstract}
Magnetic bright points are small-scale magnetic elements ubiquitous across the solar disk, with the prevailing theory suggesting that 
they form due to the process of convective collapse. Employing a unique full Stokes spectropolarimetric data set of a quiet Sun region 
close to disk centre obtained with the Swedish Solar Telescope, we look at general trends in the properties of magnetic bright points. 
In total we track 300 MBPs in the data set and we employ NICOLE inversions to ascertain various parameters for the bright points 
such as line-of-sight magnetic field strength and line-of-sight velocity, for comparison. We observe a bimodal distribution in terms of 
maximum magnetic field strength in the bright points with peaks at $\sim$480~G and $\sim$1700~G, although we cannot attribute the kilogauss 
fields in this distribution solely to the process of convective collapse. Analysis of MURaM simulations does not return the same bimodal 
distribution. However, the simulations provide strong evidence that the emergence of new flux and diffusion of this new flux play a significant role in generating the weak bright point distribution seen in our observations.
\end{abstract}

\begin{keywords}
Sun: activity -- Sun: evolution -- Sun: magnetic fields -- Sun: photosphere 
\end{keywords}




\vspace{-1.2cm}\section{Introduction}
\vspace{-0.2cm}Magnetic bright points (MBPs) were first observed in the late 1970's \citep{Dunn1973} in G-band images of the photosphere. Theoretical work \citep{Spruit79} then described the formation of such concentrated field strengths at small scales in a process termed `convective collapse'. The 
magnetic field is advected into the intergranular lanes where it concentrates. Fast downflows within the intergranular lanes coupled with pressure differences between the magnetic 
flux tubes and their surroundings results in the flux tubes `collapsing' to a point where the field strength of the tube balances pressure exerted on it from outside. In general, such balancing of magnetic field strengths with the gas pressure are called equipartition field strengths and within flux tubes these field strengths can reach values of the order of a kilogauss \citep{Leighton63, Parker1978, WebbRoberts78, Spruit79, BellotRubio2001, Nagata2008}. Optical depth unity within the flux tube is then deeper 
than the surrounding plasma and the base of the tube is then heated by the hot granular walls surrounding it. This coupled with the fact that the flux tube is partially evacuated makes it appear as a localised intensity enhancement within the intergranular lane, hence the term `magnetic bright point'. 

\citet{Nagata2008} employed {\textit{Hinode}} observations to study the formation of a single MBP, looking for the signature fast downflows within the region prior to amplification of 
the magnetic field to kilogauss strengths and the associated intensity enhancement as the MBP forms. A more rigorous statistical study using similar techniques followed and showed that the radii of the 49 studied MBPs reduced on average from 0$\arcsec$.43 to 0$\arcsec$.35 resulting in field strengths up to 1.65\,kG \citep{Fischer2009}. 

More recent studies of the magnetic field of MBPs show that it displays largely a bimodal distribution \citep{Utz2013}. In this study the authors employed {\textit{Hinode/SOT}} spectropolarimetry data to observe MBPs in four different regions (i.e, a sunspot group, near pores and in two different quiet Sun regions). Generally, the authors report a bimodal distribution with peaks around $\sim$200--300~G and $\sim$1100--1300~G across the four data sets. They were able to fit the distribution with log-normal components and suggested that the two peaks are due to the convective collapse process, with the weaker group representing uncollapsed MBPs, while the stronger group representing those that have undergone the convective collapse process, resulting in higher field strengths.

Within this work we use a unique data set with high spatial resolution spectropolarimetric information to ascertain the nature of magnetic field distributions in MBPs. The novel aspect here is that we employ MHD simulations as a comparison to determine the relation between the distribution and the process of convective collapse.

\vspace{-0.6cm}\section{Observations \& Numerical Simulations}
\begin{figure*}
\vspace{-0.2cm}
\makebox[\linewidth]{
   \includegraphics[width=0.75\linewidth]{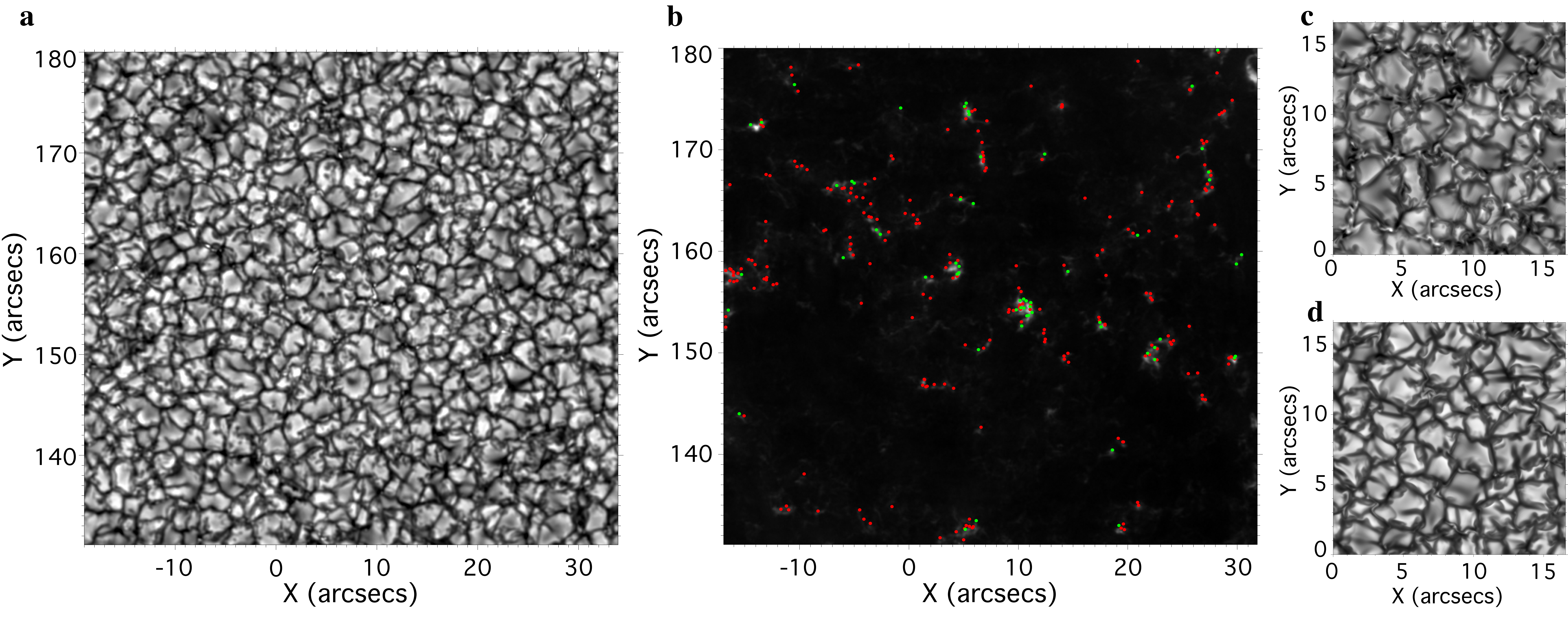}
	}
    \vspace{-0.3cm} \caption{Panel (a) shows our observations, taken in the 6301{\AA} and 6302{\AA} line pair with full Stokes spectropolarimtery CRISP scans at the Swedish Solar Telescope, of the quiet Sun at disk centre. The image is taken at 6301.0392~{\AA} in the line scan. Magnetic bright points can be seen in the intergranular lanes. Panel (b) shows the average location of the strong (\textit{green}) and weak (\textit{red}) magnetic bright points superimposed on a map of the total circular polarisation signal for the FOV averaged over the whole duration of the observations. Panel (c) and (d) show the simulated data sets we employed for the same line pair, obtained from MURaM. Panel (c) is the simulations with an initial field of 200~G while panel (d) shows an example snapshot for simulations with an initial 50~G field. The spatial sampling for the observations was 0.$''$059 pixel$^{-1}$, while the equivalent for the simulations is 0.$''$0345 pixel$^{-1}$.
}
     \label{Fig1}
\end{figure*}

\vspace{-0.2cm}The data employed in this study was acquired with the 1~m Swedish Solar Telescope \citep[SST;][]{Scharmer2003} on 2014 July 27$^{th}$ from around 14:18\,UT until 15:11\,UT of a quiet Sun region at disk centre, with an initial pointing of N0.14W4.5 in the heliographic co-ordinate system. The CRisp Imaging SpectroPolarimeter \citep[CRISP;][]{Scharmer2006, Scharmer2008} was used to 
sample the Fe~{\sc{i}} 6301{\AA} and 6302{\AA} line pair in full Stokes spectropolarimetry mode at 32 wavelength positions with a spectral FWHM of 53.5~m{\AA} and a step size of 
around 37~m{\AA} for most of the line, extending to around 77~m{\AA} at the wings. A total of 92 complete full Stokes scans were taken over the duration of the observations. The images were reconstructed 
using the Multi-Object Multi-Frame Blind Deconvolution technique \citep{vanNoort2005, Henriques2012} within the CRISPRED data reduction pipeline \citep{delaCruzRodriguez2015} 
prior to being de-rotated, aligned and destretched, which resulted in a reduced cadence for the scans of around 33~s. The effective field-of-view (FOV) of the data was approximately 
50$'' \times \, $50$''$ with a spatial sampling of 0.$''$059 pixel$^{-1}$.

The simulations were produced using the MURaM radiative MHD code \citep{Vogler2005}. This code solves large-eddy radiative three-dimensional MHD equations on a Cartesian grid, 
and employs a fourth-order Runge-Kutta scheme to advance the numerical solution in time. The numerical domain has a physical size of 12$\times$12~Mm$^2$ in the horizontal 
direction, 1.4~Mm in the vertical direction, and is resolved by 480$\times$480$\times$100 grid cells, respectively. More information on the simulations can be found in previous work \citep{Keys2011,Keys2013,Cegla2013}. Here we employ two sets of simulations, one with an initial field strength of 200~G and one with an initial field strength of 50~G. The simulations had an effective cadence of $\sim$17~s and $\sim$34~s and a duration of approximately 90~minutes and 70~minutes of physical time for the 200~G and 50~G simulations, respectively. Figure~\ref{Fig1} shows a sample image from both our observational and simulated datasets.

We employed a tracking algorithm \citep{Crockett2010} to isolate the MBPs within our observations and simulations. This algorithm is based on intensity thresholding techniques to isolate bright structures within the FOV, and has been previously used in several studies that derive on the general properties of MBPs \citep{Keys2011,Keys2014}. The application of the tracking algorithm provided 300 MBPs within the observations, 449 within the 200~G simulations and 231 within the 50~G simulations that we use for subsequent analysis.

We obtained magnetic field information from our spectropolarimetric dataset using the NICOLE inversion algorithm  \citep{SocasNavarro2015}. The inversions were run with three cycles with increasing nodes in temperature (2, 4, 7), LOS velocity (1, 2, 4) and LOS 
magnetic field (1, 2, 3) between each cycle for all MBP pixels that were tracked with the process described above. In selecting the nodes for each cycle, we used a similar approach as that described in \citet{SocasNavarro2011}. The filling factor does not vary during our inversions, but is set to 1 throughout. The initial model used was the FAL-C \citep{FALC1990} quiet Sun model. Due 
to the small signal-to-noise ratio in Stokes Q and U, which is expected as MBPs observed at disk centre are likely to be aligned vertically, we ignored Stokes Q and U in our inversions by giving them negligible weighting within the weights file for the inversion. 

To test that this is a reasonable assumption, we ran a sample of inversions including all four Stokes parameters. Again, we employed the same nodes in temperature, LOS velocity, LOS magnetic field and then increasing nodes per cycle in the horizontal magnetic field components (B$_{\rm{x}}$, B$_{\rm{y}}$: 1, 2, 3) when considering the Q and U profiles in our inversions. Across 20 randomly selected frames across the duration of the observations we sampled $\sim$500 MBP pixels for inversion of all four Stokes parameters. The retrieved inclinations in 92\% of the cases were greater than 160$^{\circ}$ (nearly perfectly vertical) with a median value of 169$^{\circ}$. These values are comparable to previous work \citep{Jafarzadeh2014}. We decided not to consider the inversion of the Q and U profiles in our subsequent analysis, as the weak signal in Q and U for our observations likely meant that the inversion code would over estimate the horizontal components \citep{Jafarzadeh2014}. Also, the weak signal meant that the fits in Q and U were poor in comparison to those we obtained in I and V. Given the inclination of the MBPs is largely vertical, however, the Q and U profiles can be omitted without issue for the total magnetic field strength, which we are interested in here.
 
\begin{figure}
\vspace{-0.2cm}
\makebox[\linewidth]{
   \includegraphics[width=0.8\linewidth]{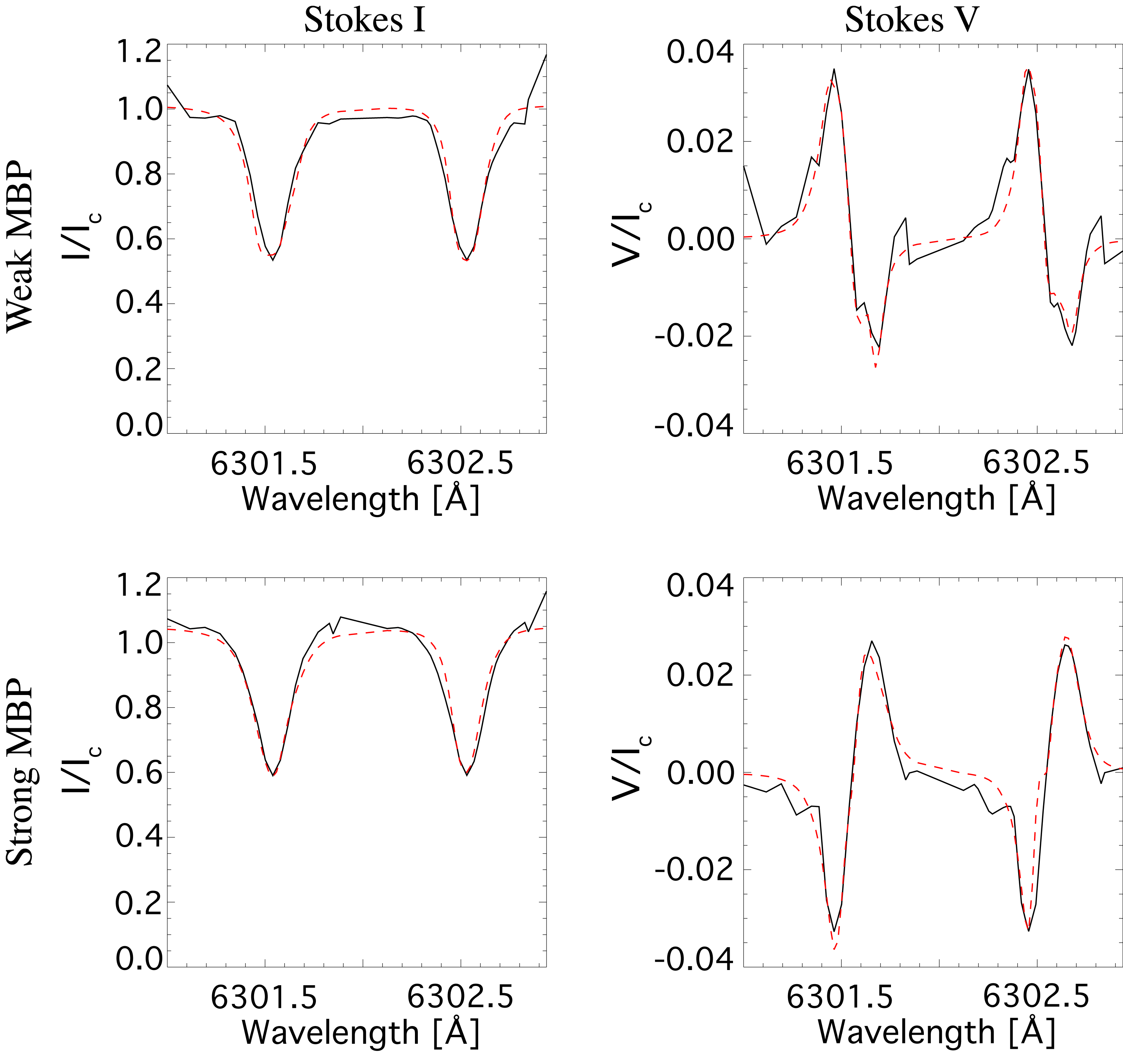}
	}
     \vspace{-0.3cm} \caption{An example of the synthetic fits obtained from the inversion of Stokes I and Stokes V observations for a weak and strong MBP. In all plots the {\textit{black}} line shows the observed Stokes I or V profile, while the {\textit{red dashed}} line shows the synthetic fit after inversion with NICOLE. The left column shows the Stokes I profile for a weak MBP (top) and a strong MBP (bottom). The right column shows the Stokes V profiles for the same two MBPs. The plots are taken for a random MBP in each distribution, and are plotted for a barycentre pixel for each. The inversions return accurate fits for our observations.
}
     \label{Fig2}
\end{figure}

Furthermore, within the weights file, more weight was specified to Stokes V in comparison to Stokes I, as the signal in Stokes V is weaker than that of Stokes I. This is a necessary step to ensure that the Stokes V profiles 
are fitted accurately by the inversion algorithm. The weights for Stokes V were altered between cycles to improve the fit of the profiles. The weights for Stokes V increased from 2 in the first two cycles to 5 in the final cycle. Examples of our synthetic fits from the inversions can be seen in Figure~\ref{Fig2}. From the models output, we were able to gather information on various properties such as LOS magnetic field, LOS velocity and gas pressure as a function of optical depth for all inverted pixels. 

\vspace{-0.6cm}\section{Results}
\vspace{-0.2cm}We examined the parameters of MBPs as a whole to search for relationships or trends. The parameters were obtained from both the MURaM outputs from our NICOLE inversions as an average value over optical depths (log $\tau$) ranging from $-1.5 $ to $ -0.5$. These general  properties are displayed in Table~\ref{Table1}. Note, that the symbol $\cdot | \div$ here denotes that the $\sigma$ boundary can be be found by multipying or dividing the mean value by the multiplicative deviation \citep[see][for more in-depth information on this definition and log-normal distributions]{Limpert2008,Utz2013}. Lifetime and intensity values are given with their respective standard deviations. We average over a range of optical depths to smooth out any inaccuracies that may fall within, or at least close to, the optical depth that the inversion code has chosen to fit the model. This removes any inconsistencies that may arise from choosing a singular optical depth. Furthermore, the range from $-1.5 $ to $ -0.5$ was chosen due to typical response functions for the line pair, which will peak closer to log $\tau$ = $-1$. Therefore, we expect the optical depth range chosen to be the region where the contribution is highest, thus, giving a more accurate representation of the MBPs. This likely corresponds to the mid-photosphere. 

\begin{table}
\centering                                    
\caption{Properties of MBPs from inversions of SST observations.}\vspace{-0.2cm}           
\label{Table1}  
\makebox[\linewidth]{    
\scalebox{0.85}{
\begin{tabular}{l c c} 
\hline\hline                        
\textbf{Property} & \textbf{Weak Group} & \textbf{Strong Group} \\
\hline       
Number of MBPs & 236 & 64 \\                            
Av. of initial B-field (G) & 510$\cdot | \div$1.6 & 840$\cdot | \div$1.6 \\ 
Av. of final B-field (G) & 530$\cdot | \div$1.5 & 900$\cdot | \div$1.6 \\
Av. of max. B-field (G) & 540$\cdot | \div$1.6 & 1700$\cdot | \div$1.2 \\
Av. of lifetime mean MBP B-field (G) & 530$\cdot | \div$1.4 & 950$\cdot | \div$1.3\\
Av. lifetime & 140$\pm$60 & 370$\pm$330\\
Av. of max. intensity (normalised units) & 1.07$\pm$0.03 & 1.10$\pm$0.04 \\
\hline                                            
\end{tabular}
}}
\end{table}

\begin{figure}
\vspace{-0.2cm}
\makebox[\linewidth]{
   \includegraphics[width=1\linewidth]{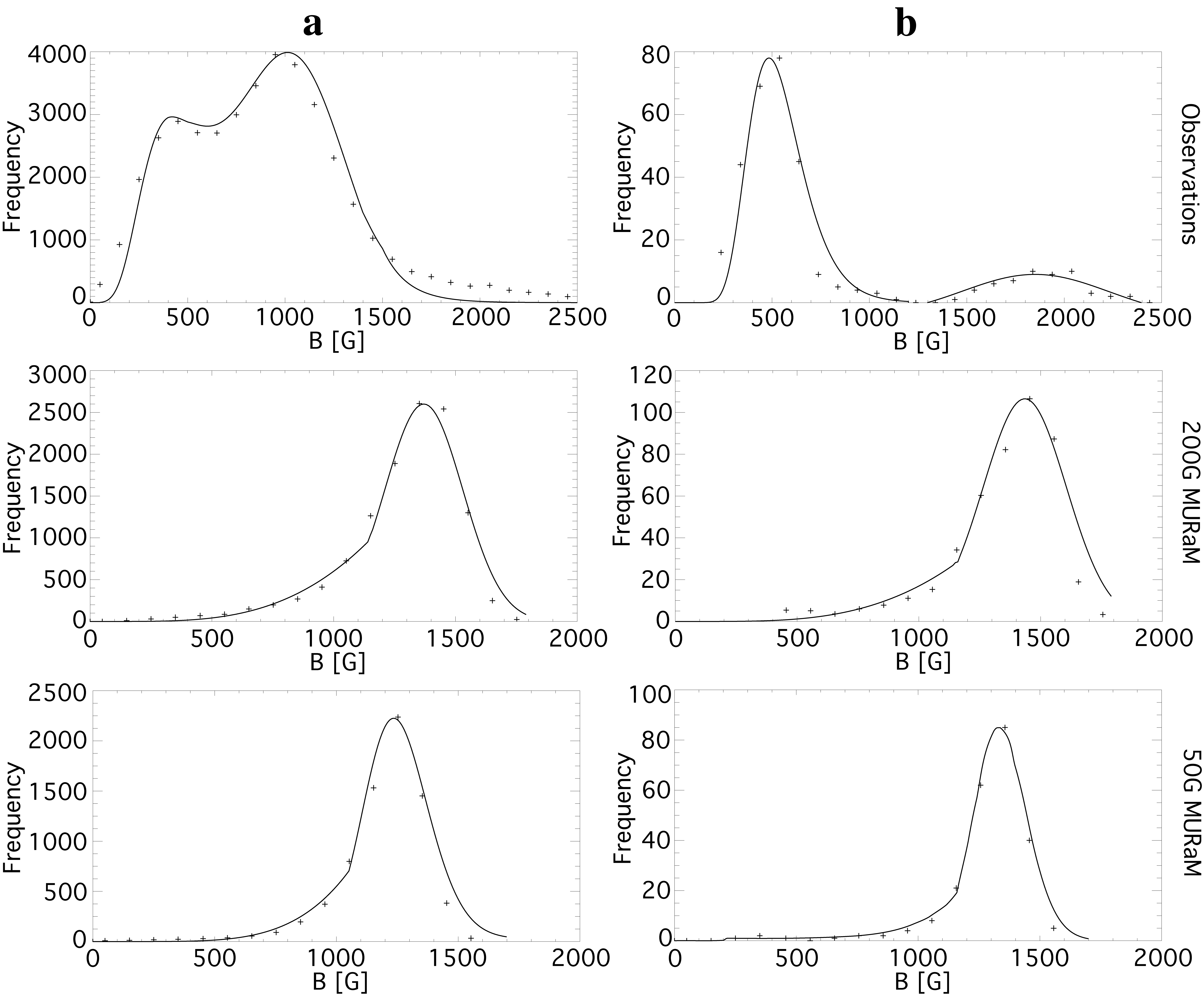}
	}
      \vspace{-0.3cm}\caption{B-field properties of the MBPs from the SST observations, 200~G MURaM simulations and 50~G MURaM simulations. Column (a) shows the B-field distribution of all tracked MBP pixels in the data. Column (b) shows the B-field distribution for the maximum B-field for each tracked MBP across their lifespan. A bimodal distribution is seen in the observations, but is not recreated in either the 200~G or 50~G simulations. Note that the distributions are displayed for the entire duration of the observations and simulations. In the case of the simulations, this represents the distribution after the field has been injected and allowed to stabilise. All distributions are fitted with a double log-normal distribution, similar to the approach adopted by \citet{Utz2013}.
}
     \label{Fig3}
\end{figure}

We investigated the general magnetic field strength distribution for tracked MBPs in two ways. We examined the maximum magnetic field strength detected in each MBP during its lifetime, and the magnetic field distribution for all detected MBP pixels within our datasets. Figure~\ref{Fig3} shows the LOS magnetic field for the observations and the simulations. Column (a) shows the LOS B-field distribution for all MBP pixels, while column (b) shows the B-field distributions for the maximum B-field found with each tracked MBP throughout their entire lifetimes. The rows show the distributions for the observations (top), the 200~G simulations (middle) and the 50~G simulations (bottom).

The B-field values derived from our spectropolarimetric inversions show two distinct groups. This distribution is similar to the one  reported in  \cite{Utz2013} and can be described with a  double peak, defined by log-normal components. The double peak nature of the distribution is more pronounced when one looks exclusively at the plot for the maximum B-field value for the 300 MBPs over their lifetime (top right plot), with the `weak' group peaking at $\sim$480~G and the `strong' group peaking at $\sim$1700~G. We note that the peak in the strong group is likely to be somewhat overestimated due to a correction factor that had to be applied to the polarimetric calibrations of all SST taken during the 2014 observing season \citep[see][for more details]{Henriques2017}. This effect is non-linear and is more pronounced at smaller B-field values, though we stress that the strong group fields will still peak above the $\sim$1300~G equipartition field strength suggested by \citet{Spruit79}. 

The bimodal distribution is less pronounced when considering all MBP pixel values. This has the effect of moving the strong peak closer (now at $\sim$1100~G) to the weak peak. This is not surprising as pixels about the barycentre of the MBP are likely to have similar LOS magnetic fields to the barycentre (where we expect the maximum peak for the MBP to be) and, therefore, will contribute more to the distribution, albeit lowering the location of the peak. Also, it is worth noting that it is expected that the magnetic field within each MBP would smoothly decline in strength from the strongest central value towards the MBP boundary, so a more continuous distribution would likely form when considering all MBP pixels.

Considering only the maximum B-field values, the average field strength of the weak group is 540\,$\cdot | \div$\,1.6~G while the strong group has an average B-field of 1700\,$\cdot | \div$\,1.2~G. 

A bimodal distribution is not apparent in MURaM simulations. The simulated datasets show a single peak that corresponds to the `strong' group in our observations with no corresponding `weak' group peak. Similar to the observations, the distribution can be fitted with a log-normal, albeit with quite a sharp tail at about 1100~G. The smaller average field strength introduced in the domain  results in a smaller number of MBPs detected in the 50~G simulations  than the 200~G simulations. The peak in the B-field distribution is slightly lower in the 50~G simulations (peaking about 1300~G as opposed to 1400~G for the 200~G simulations). This is coupled with a narrower log-normal distribution in the 50~G simulations in comparison with the 200~G simulations.  

The absence of the two-peaked distribution in the simulations may not be completely unexpected. When considering the simulations, one has to bear in mind that the simulated domain of the simulated box is small compared to the observations. Solar phenomena, such as supergranulation, are therefore missing from the simulations and that can affect bulk motion and/or the B-field distributions of MBPs. This is in part why we looked at two simulated datasets to assess if different average magnetic field strengths, that we might see between network and internetwork regions, results in the different B-field distributions. This does not seem to be the case. In fact, when we used Local Correlation Tracking \citep[LCT;][]{NovemberSimon1988} of long-duration Helioseismic and Magnetic Imager \citep[HMI;][]{Schou12} continuum images to track supergranular evolution \citep[similar to the approach employed by][]{Requerey2018} within our FOV, we do not see a correlation between the MBP positions within the network to the maximum magnetic field strength during their evolution. Panel b of Figure~\ref{Fig1} shows the spatial location of the strong and weak MBPs within the FOV.

We believe that the most likely explanation for the lack of a two peaked distribution in the simulations arises as a result of the effects of diffusion and the fact that no `new' flux is added to the simulated domain as a function of time. In terms of the diffusion within the simulated domain, the process is artificial requiring additional terms to diffuse the field. Therefore, the domain becomes more `stable' and the flux concentrations have the time to make it to kilogauss fields. 

Using the methods described in \citet{Abramenko2011}, we established the diffusion index ($\gamma$) and the diffusion coefficient ($K$) for our data and simulations. For the observations, we find values for $\gamma$ of 1.73$\pm$0.25 for the weak MBP distribution and 1.40$\pm$0.15 for the strong MBP distribution. These values suggest that the MBP motion for both weak and strong MBPs are super diffusive with similar values to other studies \citep{Keys2014}. The strong MBPs are slightly less super diffusive due to clustering in regions of higher average flux (see Figure~\ref{Fig1} panel b) and, therefore, fall within stagnation points that restrict movement. This is reflected in the values for $K$ for the two groups of 162$\pm$46~km$^2$\,s$^{-1}$ and 143$\pm$48~km$^2$\,s$^{-1}$ for the weak and strong distributions, respectively, where $K$ indicates the efficiency of the dispersion of the MBPs.

The simulations were found to have values for $\gamma$ of 1.63$\pm$0.47 and 1.77$\pm$0.54 for the 200~G and 50~G simulations, respectively. Both sets of simulations have similar diffusive properties to those of the weak group in our observations. The value for $K$ for the simulations was found to be 160$\pm$48~km$^2$\,s$^{-1}$ and 166$\pm$49~km$^2$\,s$^{-1}$ for the 200~G and 50~G simulations, respectively. This seems to be slightly higher than their strong MBP counterparts in the observations, which suggests that the diffusion within the simulations is not quite the same as those observed on the Sun, with fewer stagnation points.

Given the similarity between the simulations and the weak group in terms of diffusion, it would seem that there is another reason that accounts for the difference between the two distributions. The magnetic flux within the simulated domain remains static, in the sense that no fresh flux is added over time. In contrast, the new flux can be added to our observational field-of-view contributing to the generation of new MBPs with weaker fields which are being dispersed through turbulence prior to reaching kilogauss strengths, which is reflected in the values for $\gamma$ and $K$ that we find in our observations (i.e., the weak MBPs are more diffusive). In the simulated domain, the flux remains within the relatively small box and is able to form kilogauss concentrations over time. An examination of the weak to strong MBP number ratio in the simulations shows that it drops to 1:5 from 2:1 within 340~s in the 50~G simulations and from 4:5 to 1:5 in 357~s in the 200~G simulations. This supports the idea that the absence of new flux within the simulated domain results in the disappearance of the weak group and allows the MBPs to form kilogauss fields. This highlights the importance of flux emergence and diffusion of this new flux in MBP B-field distributions.

This double peaked distribution has sometimes been attributed to the process of convective collapse \citep{Utz2013}, with the weak field distribution attributed to uncollapsed fields. The process of convective collapse implies a correlation between the LOS B-field and the LOS downflow velocity experienced by the MBPs throughout their lifetimes. Like the LOS B-field, the LOS velocity is sampled over the same optical depth range. Response functions for the LOS velocity suggest that there are significant contributions within this height range, with more contribution between the range log $\tau$ = $0$ to $-1$ than observed for response functions of the LOS B-field. We choose the same range of optical depths as the LOS B-field, so that they are more readily comparable. We note that from tests of the variation of LOS velocity that in deeper regions the values are slightly higher in magnitude, though the trends in the rise/fall of LOS velocity are comparable between the region we sample and deeper regions. A Spearman's Rank Correlation Coefficient (SRCC) between the LOS B-field and the preceding LOS velocity for the strong MBPs in our observations gives a value of 0.202, a very weak linear trend between the two parameters. The SRCC for the weak group is 0.206, which is a similar relationship. We note that preceding here refers to a time frame of 297~s before (and including) the peak B-field. We include the frame of peak B-field here in case the peaks in both parameters are coincident due to the temporal resolution. These values suggest that there is no real correlation between the magnetic field strength and the preceding LOS velocity, or it is very weak at best. It should be noted as well, however, that it cannot be discounted that convective collapse could occur on a timescale shorter than our scan time of 33~s, although a large number of studies of convective collapse have comparable temporal resolution and find timescales for collapse on the order of minutes \citep{Nagata2008, Fischer2009}, which would be sufficiently sampled by our temporal resolution. Furthermore, it should be noted that counter streaming flows \citep[as observed by][for example]{BellotRubio2001, Utz2014} could further complicate any treatment of LOS velocities, although by considering preceding values, we expect to miss the rebound shocks described by \citet{BellotRubio2001}. 

Similar values were observed when applied to our simulations (SRCC = 0.115 for 200~G and -0.239 for 50~G). These values suggest that there is a very weak (at best) linear trend between the magnetic field and the preceding LOS velocity. Now, this does not necessarily suggest that convective collapse does not occur, only that the strength of the B-field is not reliant on the strength of the preceding downflow. Given the fact that strong downflows are required for collapse to occur, these values suggest that convective collapse is not solely responsible for the stronger B-fields in the MBPs and, therefore, it is not solely responsible for the bimodal distribution of magnetic field strengths as was once thought. Work by \citet{Utz2014} comes to similar conclusions, in that it seems that other processes besides convective collapse can result in kilogauss fields in MBPs. Within this work the authors study a large sample of MBPs and find evidence for kilogauss fields without the telltale signatures of convective collapse. The results from our study of the simulations suggests that flux emergence and diffusion of new flux plays a significant role in this distribution and the ability for MBPs to form kilogauss structures.

\vspace{-0.8cm}\section{Conclusions}
\vspace{-0.2cm}In this study we analysed high resolution spectropolarimetric data of a quiet Sun region at disk centre to examine the general magnetic properties of magnetic bright points (MBPs). We find a bimodal distribution of magnetic field strengths in MBPs. We do not find significant evidence that this distribution results from the process of convective collapse in MBPs. Furthermore, the spatial location of MBPs within network cells do not seem to explain this bimodal distribution in terms of B-field strength, i.e., the distribution does not appear to arise based on whether the MBPs are classed as network or internetwork MBPs.

We examined MURaM simulations to search for evidence of the bimodal distribution in simulated datasets. We investigated two sets of simulations, one with an initial field of 200~G and one with an initial field of 50~G. However, neither return the bimodal distribution that we see in the observations. On closer inspection, we determined that the simulated MBPs had similar diffusive properties as the weak MBP distribution in our observations. However, on examining the ratio of weak to strong MBPs over time in the simulations, we see that the weak distribution rapidly declines with the number of strong MBPs increasing. We attribute this to the fact that no new flux is added to the simulated domain, so the magnetic field has more opportunity to amplify to kilogauss field strengths. Within the observations, new flux is constantly added, which disperses prior to reaching kilogauss fields, thus, producing the bimodal distribution we observe.

The evidence that we find in our observations and simulations suggests that this bimodal distribution is not solely the result of the convective collapse process. We see that diffusion and emergence of new flux plays a significant role in the distribution of magnetic fields at these scales on the solar surface.


\vspace{-0.65cm}\section*{Acknowledgements}
\vspace{-0.15cm}
All authors are grateful to the anonymous referee whose comments enhance the manuscript. P.H.K. is grateful to the Leverhulme Trust for the award of an Early Career Fellowship. 
P.H.K would like to thank Drs. David Kuridze and Hector Socas-Navarro for assistance in setting up the inversion codes. V.M.J.H. was supported by the Research Council of Norway, 
project number 250810. Both V.M.J.H. and S.J. were supported by the Research Council of Norway through its Centres of Excellence scheme, project number 262622. S.J. also 
acknowledges support from the European Research Council (ERC) under the European Union's Horizon 2020 research and innovation program (grant agreement No. 682462). D.B.J. 
wishes to thank the UK Science and Technology Facilities Council (STFC) for the award of an Ernest Rutherford Fellowship alongside a dedicated Research Grant. D.B.J. also wishes to 
thank Invest NI and Randox Laboratories Ltd. for the award of a Research \& Development Grant (059RDEN-1) that allowed this work to be undertaken. This research was undertaken 
with the assistance of resources and services from the National Computational Infrastructure (NCI), which is supported by the Australian Government.  This work used the DiRAC Data 
Centric system at Durham University, operated by the Institute for Computational Cosmology on behalf of the STFC DiRAC HPC Facility. DiRAC is part of the UK National E-Infrastructure.
Observations were acquired within the SolarNet Project Transnational Access Scheme. SolarNet is a project supported by the EU-FP7 under Grant Agreement 312495. The SST is 
operated on the island of La Palma by the Institute for Solar Physics of Stockholm University in the Spanish Observatorio del Roque de los Muchachos of the Instituto de 
Astrof{\'{i}}sica de Canarias. We acknowledge support from the STFC Consolidated Grant to Queen's University Belfast.



\label{lastpage}

\end{document}